\documentclass[openacc]{rstransa}

\jname{rsta}
\Journal{Phil. Trans. R. Soc}

\titlehead{Research}

\newcommand{\aap}{    {Astron. Astrophys. }}

\newcommand{\apj}{    {Astrophys. J. }}
\newcommand{\apjl}{   {Astrophys. J. Lett. }}
\newcommand{\apjs}{   {Astrophys. J. Suppl. }}

\newcommand{\mnras}{  {Mon. Not. Roy. Astron. Soc. }}

\newcommand{\nat}{    {Nature }}

\newcommand{\pre}{    {Phys. Rev. E }}
\newcommand{\prl}{    {Phys. Rev. Lett. }}
\newcommand{\solphys}{{Solar Phys. }}
 
\newcommand{\ssr}{    {Space Sci. Rev. }}

\begin{document}
\title{Detector for fast wave trains in the solar radio emission} 
\author{V. A. Dmitriev$^{1,2}$, E. G. Kupriyanova$^{1,2,3}$,\\
A. V. Mikhalchuk$^{1}$}

\address{$^{1}$Saint Petersburg State University, 199034 Saint Petersburg, Russia\\
$^{2}$Central Astronomical Observatory at Pulkovo of RAS, 196140 Saint Petersburg, Russia\\
$^{3}$ Ioffe Physical-Technical Institute of RAS, 194021 Saint Petersburg, Russia}
\subject{Solar physics}
\corres{V. A. Dmitriev, E. G. Kupriyanova
\email{st069138@student.spbu.ru},\\
\email{elenku@bk.ru}}

\begin{abstract}
Quasi-periodic fast propagating (QFP) wave trains observed in the solar corona after some energetic events (solar flares, coronal mass ejections, jets) open possibilities for diagnostics of spatial and temporal scales of the impulsive energy release processes, that are absent in the standard model of a solar flare. 
Besides, the dynamics of the wave trains and their characteristic spatial and temporal signatures allow to localize the initial energy release volume magenta and to perform fine diagnostics of the transverse structures of plasma inhomogeneities in the solar corona.
However, the small number of such events registered significantly limits their promising diagnostic potential. 
The aim of this paper is to perform an automatic search for fast wave trains in radio data.
We apply classifying neural network/machine learning methods. Dynamic radio spectra obtained by HiRAS radio spectrographs within the 20 MHz -- 2.5 GHz frequency band during  2011 were used. We consider 50 global coronal EUV waves as marker events for more a targeted search in HiRAS data.
Our automatic detector revealed 50 independent QFP-candidates events with the temporal signatures similar to that of the fast wave trains, with 13 candidates connected with the global waves.
\end{abstract}

\maketitle

\section{Introduction}
The study of active processes in the solar atmosphere, such as solar flares and coronal mass ejections (CMEs)
remains one of the central research topics for the solar physics community as these events influence the near-Earth space environment and space-borne and ground-based infrastructures.

According to the standard solar flare model, the magnetic energy  accumulated in an active region is released by magnetic reconnection (see, e.~g., \cite{Hudson2016SoPh..291.1273H} and references therein). 
However, the flare energy release rate and locations and scales of the solar flare volume are still debated (see, e.~g., \cite{Kerr2023BAAS...55c.206K, Russell2025arXiv251211631R, Druett2025arXiv251224479D} for the review). 

The theory solves the problem of the reconnection rate, which appear to be too slow in comparison with the typical observed dynamics of solar flare emission in the Sweet-Parker model  \cite{Sweet1958IAUS....6..499S, Sweet1958NCim....8S.188S, Parker1963ApJS....8..177P}, by considering magnetohydrodynamic (MHD) shocks \cite{Petschek1964NASSP..50..425P}, tearing-mode instability \cite{Shibata2001EP&S...53..473S, Pucci2014ApJ...780L..19P}, or the presence of the stochastic magnetic component \cite{Lazarian1999ApJ...517..700L} in the current sheet.  
Another mechanism is oscillatory reconnection, which is able to produce, in addition to the fast reconnection rate, quasi-periodic pulsations during solar flares \cite{McLaughlin2018SSRv..214...45M}.

There are indirect methods of determining the scales of flare energy releases.
Temporal scales of the energy release  can be inferred from the evolution of the impulsive phase in microwave and hard X–ray (HXR) bands.
An increase in emission in these bands indicates 

a rise in the number of non-thermal electrons accelerated during magnetic reconnection.
The dynamics of the impulsive phase, once  both the acceleration and emission mechanisms are properly taken into account, therefore, provides an estimation of the rate of energy release. But the acceleration process could be suppressed in the strong magnetic field \cite{Arnold2021PhRvL.126m5101A}.

Determining the spatial extent of the energy release region, remaining one of the principal parameters for estimating the solar flare energetics, is considerably more uncertain. 
In the microwave or HXR images one may estimate the flare source size by measuring the extent of the emission above the apex of flare-loop coronal arcades at the solar limb and limiting it with an isoline 
at a chosen brightness level \cite{Masuda1994Natur.371..495M, Gary2018ApJ...863...83G}.
However, it should be noted that such estimates strongly depend on the subjective selection of the isoline level, on the instrumental parameters (sensitivity, spatial resolution), and on the underlying emission mechanism. 

Another observational challenge is the localization of the energy release region in the solar atmosphere.
Modern instruments and spatial spectropolarimetric techniques can,
in some cases, identify the electron-acceleration region \cite{Gary2023BAAS...55c.125G}.
However, the authors consistently find that the acceleration region does not coincide with the reconnection site,
indicating the need for independent diagnostics of both the scales and the location of the energy release.

A promising approach on this way is the analysis of Quasi-periodic Fast Propagating (QFP) wave trains in the solar atmosphere \cite{Roberts1984ApJ...279..857R,Liu2011ApJ...736L..13L}. 
The trigger of the QFP wave train is a localized impulsive energy release
while the plasma inhomogeneities in the corona (coronal loops, holes, current sheets, etc.) act as waveguides for MHD waves.
The propagation speeds indicate that QFP waves are fast MHD modes (e.~g., \cite{Shi2025ApJ...990....1S, Shi2026ApJ...996...72S}). 
However, dispersion over the both launch angle and phase speed of the initial perturbation together with the presence of the waveguide lead over time to an interference pattern~--- a wave train traveling within the waveguide along the magnetic field lines (the \emph{trapped} mode, \cite{Meszarosova2014ApJ...788...44M}).
The remaining part of the perturbation leaks out of the guide (the \emph{leaky} mode) \cite{Nistico2014A&A...569A..12N, Pascoe2013A&A...560A..97P}.
Both trapped and leaky components may be produced by the same trigger \cite{Nistico2014A&A...569A..12N, Shi2025ApJ...990....1S}.
As the wave train propagates, it modulates parameters of the medium and, consequently, the radio emission produced in that medium, both inside and outside the guide. The uniqueness of the diagnostic potential of QFP wave trains is that the characteristic shape of the wave train and, therefore, of its wavelet spectrum depends on temporal and spatial parameters of the trigger,

the distance between the detection point and the trigger, and the transverse waveguide structure \cite{Nakariakov2024RvMPP...8...19N}.

Unfortunately, this potential remains largely untapped, mainly because of only a few dozen events which have been reported to date as case studies.
The principal signature of wave trains is the characteristic shape (\emph{tadpole}) of a wavelet spectrum that is a combination of a narrow-band tail and a broad-band head \cite{Nakariakov2004MNRAS.349..705N}
or a more complicated \emph{boomerang} shape \cite{Kolotkov2021MNRAS.505.3505K}.

The tadpole-shaped wave train was first identified during a solar eclipse in the Fe\,{\sc xiv} 5303\,\AA\ line \cite{Katsiyannis2003A&A...406..709K}.

After that, similar signatures have been identified in radio emission at frequencies below 4.5\,GHz
(gyro-synchrotron and plasma mechanisms) \cite{Meszarosova2009ApJ...697L.108M}. 
The launch of the Solar Dynamics Observatory (SDO) provided the capability to detect these signatures in extreme ultraviolet (EUV) difference images and to follow their spatiotemporal evolution \cite{Liu2011ApJ...736L..13L}. 

Since these pioneer papers, a few dozen events detected in EUV \cite{Shen2022SoPh..297...20S} and radio \cite{Meszarosova2009A&A...502L..13M, Karlicky2011A&A...529A..96K, Karlicky2013A&A...550A...1K} have been reported. A statistically meaningful study requires a much larger sample. A related question that remains open is whether wave trains are genuinely that rare. In principle, each impulsive energy release should trigger the wave train. Therefore, the goal of our work 

is to conduct a systematic search for QFP waves in radio data and to assemble them into a catalog.

In this paper, we present a prototype automatic detection system (the detector, hereafter) for QFP wave trains 
in solar dynamic radio spectra.
The detector is based on a binary classification neural network that has been trained and validated on synthetic flare time profiles.
In Section~\ref{sec:method} we describe the methodology and report results of tests on synthetic data. 
Section~\ref{sec:radiodata} explains how flare time profiles data are selected and processed.

Section~\ref{sec:results} summarizes the findings of the search for wave trains in the selected radio data.
Finally, in Section~\ref{sec:discussion} we discuss the implications of our results and outline the prospects for creating a large scale catalog of QFP wave trains.

\section{Methodology}
\label{sec:method}

In many published works, the identification of QFP wave trains was based on the characteristic shape of their wavelet spectrum, or cross-correlations or their combination \cite{Katsiyannis2003A&A...406..709K, Meszarosova2009A&A...502L..13M, Meszarosova2009ApJ...697L.108M}.

However, this approach is not well-suited for automated detection. The required spectral shape is not clearly defined, and 
other phenomena, like red noise, can produce a similar spectrum.

Therefore, we have developed an automated detection scheme based on the classification of the flare time profile adapting the approach developed by \cite{Belov2024ApJS..274...31B} for finding quasi-periodic damping oscillations with constant period.
The core of both our approach and approach  of \cite{Belov2024ApJS..274...31B} uses a 1D convolutional neural network (CNN) \cite{WangCNN2017}.
We have  added average pooling layers after every convolution block and have used a fixed input length size for fully convolutional network (FCN) (see Figure~1 (b) in \cite{WangCNN2017}).
In our scheme, CNN inputs a detrended flare time profile at fixed size and performs binary classification as either does it contains a wave train or does not.

To perform supervised learning for a CNN, a large dataset of QFP samples is required. Unfortunately, the number of wave train registrations in solar data~--- currently around several dozen cases~--- is insufficient for this purpose. Therefore, in order to train and test the neural network, we generate synthetic solar flare time profiles containing wave trains in sufficient quantities.

\subsection{Synthetic flare time profiles}
\label{sec:synthetic}
The length of each synthetic time series is fixed at $N = 320$ counts.
To construct a synthetic time profile close to that of a real solar flare, we combine three components: (1) a slowly varying trend, which simulates the general dynamics of the flare emission; (2) a quasi-periodic component, representing the sought wave train; and (3) noise, including both instrumental and natural noise caused by, for example, inhomogeneities in the emission source or effects of emission transfer from the source to the observer. These three components can be combined in different ways. If, for example, a quasi-periodic signal depends on or is defined by the flare energy release process~--- the stronger the emission, the higher the amplitude of oscillations~--- then a multiplicative model could be used \cite{Kupriyanova2013SoPh..284..559K}. To prevent the flare trend from distorting the QFP signal, we adopt a simple additive model for the synthetic time series as a starting point \cite{Broomhall2019ApJS..244...44B}:
\begin{equation}\label{eq:s}
S(t) = T(t) + R(t) + N(t).
\end{equation}
For the trend $T(t)$ we use the averaged models of the simplest solar and stellar flares:
polynomial on rise and exponential on decay trend \cite{Davenport2014ApJ...797..122D, Kashapova2021SoPh..296..185K};
polynomial \cite{Gryciuk2017SoPh..292...77G}, and asymmetric gaussian function \cite{Broomhall2019ApJS..244...44B}.

The signal is defined as one of three possible component
$$
R(t)=
\begin{cases}
Q(t), & \text{50\% of all signals;}\\
B(t), & \text{20\% of all signals;}\\
0,    & \text{30\% of all signals;}
\end{cases}
$$
where $Q(t)$ is the "true" signal (the wave train), $B(t)$ is the "bad" signal (single short spike, sinusoid, or beat), and $0$ indicates the absence of a signal.
The noise $N(t)$ is a superposition of the white noise and the red noise. The white noise is always present, the red noise appears in 75\% of the profiles. 
The red noise correlation coefficient is log-uniform distributed $(0.75, 0.99)$.
The relative white/red ratio is log-uniform distributed in the range $(1, 2)$.
The overall signal-to-noise ratio (SNR) is log-uniform distributed $(1, 2)$.
Note that the real data have a low SNR rate so using the log-uniform distribution we increase a percentage of time profiles with the lower SNR in comparison with the uniform distribution and reduce SNR bias difference between real and synthetic datasets.
Full generation parameters of trends are presented in table~\ref{tab:trend}.

We should clarify the choice of the SNR distribution. We define the SNR as the ratio of $\sigma_Q / \sigma_N$, where $\sigma_N$ is the standard deviation of the noise component and $\sigma_Q$ is the standard deviation of a wave train after robust energy normalization (excluding high peaks and low amplitude counts). The reason for this normalization is that the different QFP wave trains can have different numbers of high-amplitude oscillation cycles. An example of this difference one can find in the upper left panel of Figure 2 in
\cite{Kolotkov2021MNRAS.505.3505K}. This difference can lead to an underestimation of the standard deviation for the left time profile in that panel, and to its overestimation for the right profile. By introducing energy normalization, we make the simulated time profiles more similar to real data, which is usually very noisy, and where the low-amplitude parts of the wave trains would be unrecognizable due to the noise. Using the above definition of SNR, we found that SNR > 2 is extremely high and unlikely to exist in real data.

We should mention that the existing numerical simulations obtain the characteristic form of the wave train for the phase speed \cite{Kolotkov2021MNRAS.505.3505K} or for the plasma density \cite{Nakariakov2004MNRAS.349..705N}. But we develop the detector for the wave trains in, finally, different types of flare emission. The dependence of the intensity on the plasma density is quadratic in the EUV band, i.~e., for the thermal emission, but, it could be linearised \cite{Cooper2003A&A...409..325C}. 
For the gyrosynchrotron emission (usually, at radio frequencies $f > 3$~GHz for the solar corona), this dependence is more complicated because of, for example, Razin effect at the lower frequencies in the optically thick part of the gyrosynchrotron spectrum \cite{Melnikov2008SoPh..253...43M, Kuznetsov2012Ge&Ae..52..883K}.
Besides, the plasma density variations lead to variation of the magnetic field. The intensity of
the gyrosynchrotron emission depends nonlinearly both on the magnetic field strength and on the angle between the magnetic field and the line-of-sight \cite[for the simplified expressions]{Dulk1982ApJ...259..350D}. So, small variation of the magnetic field can result in significant variations of the gyrosynchrotron intensity. For optically thin large-amplitudes the coherent plasma emission (usually, for $f < 2$~GHz), this dependence is also non-linear. The non-linearity would modulate radio frequency \cite{Kuznetsov2006SoPh..237..153K} or spectrum \cite{Kudryavtsev2021MNRAS.503.5740K} of the plasma emission. However, both for the gyrosynchrotron and plasma emissions, the characteristic time scales distribution of the intensity variations are not affected.

For the basic wave train $Q_0(t)$ we used the time profiles of wave trains obtained from MHD simulations, as well as the time profiles extracted from radio observations after wavelet filtering \cite{Meszarosova2009ApJ...697L.108M}.
Time‑profiles were selected from the following works:
\cite{Nakariakov2004MNRAS.349..705N} (Figures 3–5),
\cite{Nakariakov2005SSRv..121..115N} (Figure 2),
\cite{Meszarosova2009ApJ...697L.108M} (Figure 2),
\cite{Meszarosova2014ApJ...788...44M} (Figures 3, 5, 7),
\cite{Goddard2019A&A...624L...4G} (Figures 2, 3),
\cite{Kolotkov2021MNRAS.505.3505K} (Figures 2, 3, 5).
Only time profiles with a single branch in the wavelet spectrum ("tadpole" type) were chosen; more complex ("boomerang" type) profiles were excluded.
The curves were digitized, resulting in 38 basic profiles $Q_0(t)$.
The temporal scale distribution of QFP currently is unknown.
Therefore all $Q_0(t)$ are temporal normalized in the time interval $[10;90]$.

\begin{figure}[ht]
    \centering
    \includegraphics[width=\linewidth]{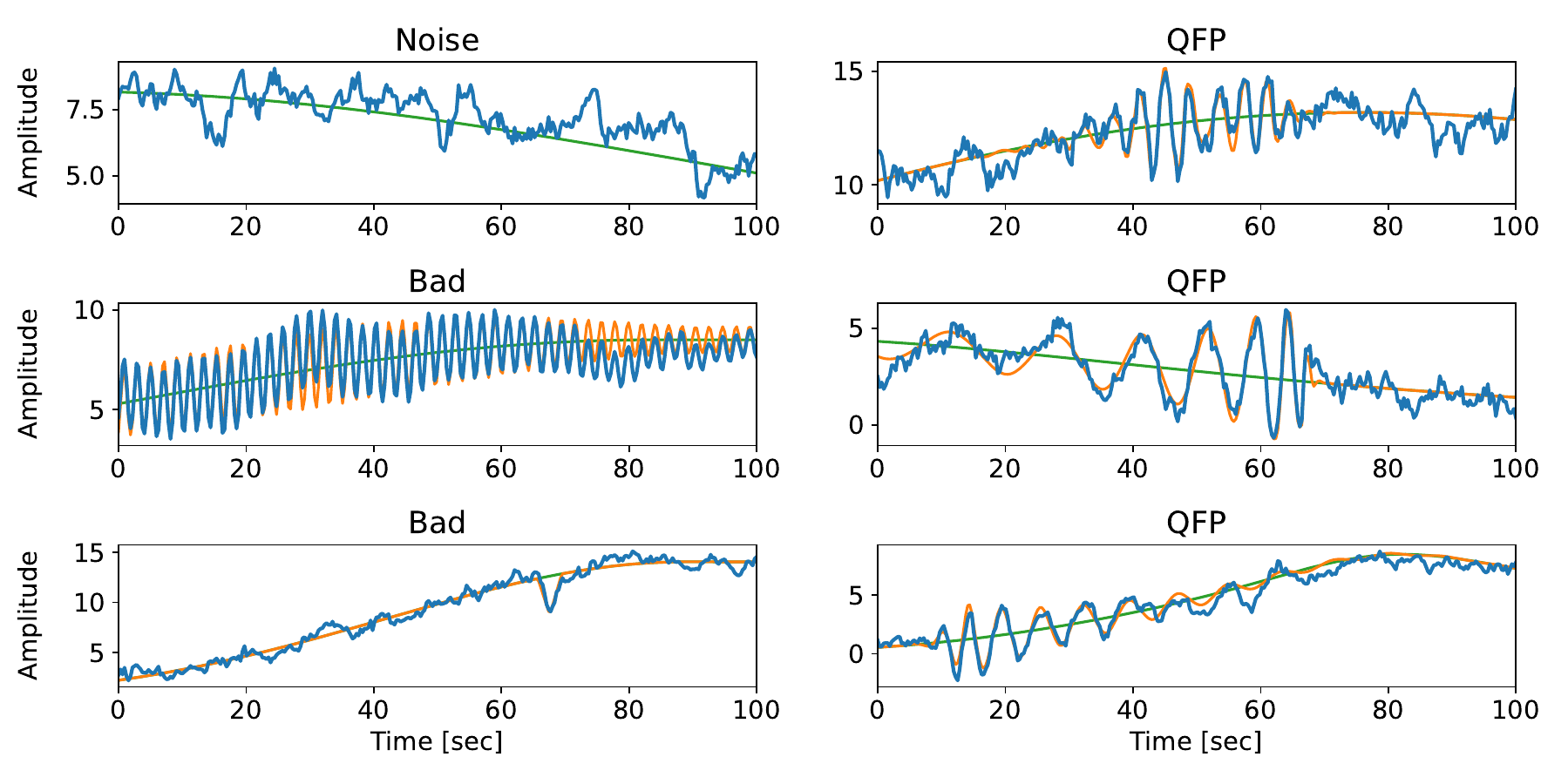}
    \caption{Example of generated synthetic time profiles. The green curve is trend $T(t)$. The orange curve is $T(t) + R(t)$. The blue curve is full signal with noise $S(t)$.}
    \label{fig:synth}
\end{figure}

However, for supervised learning a neural network a much larger diversity and number of samples is required, so we performed augmentation of the basic time profiles by scaling and shifting the time profiles.
The characteristic rise and decay times were also varied by adjusting the parameters $\sigma_{\text{rise}}$ and $\sigma_{\text{decay}}$ as follows:
\begin{equation}\label{eq:q}
Q(t) = Q_0 \left( \tau + t_{\text{peak}}\right) \cdot
\left\{
\begin{alignedat}{3}
    & \exp \left( - \frac{\tau^2}{2 \sigma_{\text{rise}}^2}  \right), & \quad \tau    < 0, \\
    & \exp \left( - \frac{\tau^2}{2 \sigma_{\text{decay}}^2} \right), & \quad \tau \geq 0; \\
\end{alignedat}
\right.
\qquad \qquad
\tau = \frac{t - t_{\text{shift}} - t_{\text{peak}}}{t_{\text{scale}}}.
\end{equation}
Here $t_{\text{shift}}$ is the time shift,
$t_{\text{peak}}$ is the time of maximum amplitude of QFP,
$t_{\text{scale}}$ is the temporal scaling parameter (expansion if $t_{\text{scale}} > 1$, compression if $t_{\text{scale}} < 1$).
The full set of generated parameters of $Q(t)$ are presented in Table~\ref{tab:qfp}.
As a consequence of the augmentation process, a dataset of $10^6$ synthetic time profiles was obtained including also "bad" and noisy signals.
Six samples are selected from this set and shown in Figure~\ref{fig:synth}.

\begin{table}[t]
\centering
\begin{tabular}{l|c|c|c|c}
Trend type & Parameters & Position & $t_{\text{shift}}$ & $t_{\text{scale}}$ \\ \hline
           & $B \sim U(0.0, 5.0)$ & rise     & $(75.0, 95.0)$ & $(30.0, 50.0)$ \\
Gryciuk    & $C \sim U(1.0, 5.0)$ & full     & $(5.0, 25.0) $ & $(5.0, 30.0) $ \\
           & $D \sim U(0.5, 1.0)$ & decay    & $(-20.0, 5.0)$ & $(30.0, 60.0)$ \\ \hline

           & $\sigma_{\text{rise}}  \sim U(1,3)$  & rise     & $(75.0, 90.0)$  & $(10, 50)$    \\
Broomhall  & $\sigma_{\text{decay}} \sim U(5,15)$ & full     & $(25.0, 45.0)$  & $(2.0, 5.0)$  \\
           &                                      & decay    & $(-20.0, 5.0)$  & $(5.0, 20.0)$ \\ \hline

           &     & rise     & $(75.0, 95.0) $ & $(30.0, 160.0)$ \\
Davenport  & --- & full     & $(15.0, 35.0) $ & $(15.0, 50.0)$  \\
           &     & decay    & $(-20.0, 10.0)$ & $(50.0, 300.0)$ \\
\end{tabular}
\caption{Probabilities of choosing every trend type and position are equal.
$t_{\text{start}}$ is uniformly distributed; $t_{\text{scale}}$ is log-uniformly distributed.
The amplitude (maximum value) is normal distributed $\sim \mathcal{N}(10,3)$, negative values converted to zero.}
\label{tab:trend}
\end{table}

\begin{table}
\centering
\begin{tabular}{l|c|c|c|c|c|c}
QFP type source & $W$ & $t_{\text{peak}}$ & $t_{\text{shift}}$ & $t_{\text{scale}}$ & $\sigma_{\text{rise}}$ & $\sigma_{\text{decay}}$ \\ \hline
\cite{Kolotkov2021MNRAS.505.3505K}, Figure 2, $z_0 = 20$  &   1 & 78.027 & $(-40.0, 10.0)$ & $( 1.0, 2.2)$ & $(20.0, 150.0)$ & $( 2.0,  10.0)$ \\
\cite{Kolotkov2021MNRAS.505.3505K}, Fig. 2, $z_0 = 60$  &   1 & 82.570 & $(-40.0, 10.0)$ & $( 0.8, 3.0)$ & $(10.0, 150.0)$ & $( 0.7,  10.0)$ \\
\cite{Kolotkov2021MNRAS.505.3505K}, Fig. 2, $z_0 = 135$ &   1 & 79.550 & $(-50.0, 10.0)$ & $( 1.2, 6.0)$ & $( 3.0,  70.0)$ & $( 0.6,  10.0)$ \\
\cite{Kolotkov2021MNRAS.505.3505K}, Fig. 3, $p0 = 1.0$  &   1 & 57.484 & $(-20.0, 30.0)$ & $( 1.0, 5.0)$ & $( 5.0, 100.0)$ & $( 0.6,  10.0)$ \\
\cite{Kolotkov2021MNRAS.505.3505K}, Fig. 5, $z0 = 150$  &   1 & 75.030 & $(-25.0, 15.0)$ & $( 1.2, 4.0)$ & $( 7.0, 100.0)$ & $( 0.4,   7.0)$ \\
\cite{Nakariakov2004MNRAS.349..705N}, Fig. 3      &   1 & 76.288 & $(-35.0,  5.0)$ & $( 0.5, 1.5)$ & $(15.0, 150.0)$ & $( 2.0,  30.0)$ \\
\cite{Nakariakov2004MNRAS.349..705N}, Fig. 4      &   1 & 66.920 & $(-25.0, 10.0)$ & $( 0.5, 1.5)$ & $(15.0, 150.0)$ & $( 3.0,  30.0)$ \\
\cite{Nakariakov2004MNRAS.349..705N}, Fig. 5      &   1 & 50.546 & $(-15.0, 25.0)$ & $( 0.4, 1.5)$ & $(15.0, 150.0)$ & $( 2.5,  25.0)$ \\
\cite{Nakariakov2005SSRv..121..115N}, Fig. 2, a   &   1 & 54.253 & $(-20.0, 25.0)$ & $( 0.6, 1.5)$ & $(10.0, 150.0)$ & $( 2.0,  25.0)$ \\
\cite{Nakariakov2005SSRv..121..115N}, Fig. 2, b   &   1 & 49.685 & $(-20.0, 25.0)$ & $( 0.5, 1.5)$ & $(20.0, 150.0)$ & $( 2.0,  25.0)$ \\
\cite{Nakariakov2005SSRv..121..115N}, Fig. 2, c   &   1 & 46.966 & $(-20.0, 30.0)$ & $( 0.5, 1.5)$ & $(15.0, 150.0)$ & $( 2.0,  25.0)$ \\
\cite{Nakariakov2005SSRv..121..115N}, Fig. 2, d   &   1 & 45.113 & $(-20.0, 30.0)$ & $( 0.5, 1.5)$ & $(10.0, 100.0)$ & $( 2.0,  25.0)$ \\
\cite{Meszarosova2009ApJ...697L.108M}, Fig 2, 1   &   1 & 55.555 & $(-10.0, 15.0)$ & $(0.25, 1.1)$ & $(15.0, 200.0)$ & $(10.0, 100.0)$ \\
\cite{Meszarosova2009ApJ...697L.108M}, Fig 2, 2   &   1 & 46.320 & $(-10.0, 20.0)$ & $(0.25, 1.1)$ & $(20.0, 200.0)$ & $( 5.0, 100.0)$ \\
\cite{Meszarosova2009ApJ...697L.108M}, Fig 2, 3   &   1 & 45.450 & $(-10.0, 25.0)$ & $(0.25, 1.1)$ & $(20.0, 200.0)$ & $( 5.0, 100.0)$ \\
\cite{Meszarosova2014ApJ...788...44M}, Fig. 7, a  & 1.3 & 47.213 & $(-15.0, 10.0)$ & $( 0.5, 1.5)$ & $(10.0, 100.0)$ & $(20.0, 100.0)$ \\
\cite{Meszarosova2014ApJ...788...44M}, Fig. 7, b  & 1.3 & 60.794 & $(-25.0, 10.0)$ & $( 0.5, 1.5)$ & $(10.0, 100.0)$ & $(20.0, 100.0)$ \\
\cite{Meszarosova2014ApJ...788...44M}, Fig. 7, c  & 1.3 & 58.238 & $(-20.0, 10.0)$ & $( 0.5, 1.5)$ & $(10.0, 100.0)$ & $(15.0, 100.0)$ \\
\cite{Meszarosova2014ApJ...788...44M}, Fig. 7, d  & 1.3 & 53.888 & $(-10.0, 10.0)$ & $( 0.5, 1.5)$ & $(10.0, 100.0)$ & $(20.0, 100.0)$ \\
\cite{Meszarosova2014ApJ...788...44M}, Fig. 3, a1 &   1 & 55.303 & $(-20.0, 15.0)$ & $( 0.4, 1.5)$ & $(10.0, 200.0)$ & $( 4.0,  50.0)$ \\
\cite{Meszarosova2014ApJ...788...44M}, Fig. 3, a2 &   1 & 57.180 & $(-15.0, 15.0)$ & $( 0.4, 2.0)$ & $( 6.0, 150.0)$ & $( 2.5,  50.0)$ \\
\cite{Meszarosova2014ApJ...788...44M}, Fig. 3, a3 &   1 & 58.896 & $(-15.0, 10.0)$ & $( 0.5, 1.5)$ & $( 5.0,  50.0)$ & $( 2.5,  50.0)$ \\
\cite{Meszarosova2014ApJ...788...44M}, Fig. 3, b1 &   1 & 57.052 & $(-25.0, 15.0)$ & $( 0.4, 1.2)$ & $(15.0, 100.0)$ & $( 5.0,  50.0)$ \\
\cite{Meszarosova2014ApJ...788...44M}, Fig. 3, b2 &   1 & 62.967 & $(-25.0, 10.0)$ & $( 0.5, 2.0)$ & $(15.0, 100.0)$ & $( 2.5,  50.0)$ \\
\cite{Meszarosova2014ApJ...788...44M}, Fig. 3, b3 &   1 & 73.683 & $(-20.0, 10.0)$ & $( 0.5, 2.0)$ & $(10.0, 100.0)$ & $( 1.5,  15.0)$ \\
\cite{Meszarosova2014ApJ...788...44M}, Fig. 3, d2 &   1 & 52.284 & $(-15.0, 15.0)$ & $( 0.5, 2.0)$ & $(10.0, 100.0)$ & $( 2.0,  50.0)$ \\
\cite{Meszarosova2014ApJ...788...44M}, Fig. 3, d3 &   1 & 59.770 & $(-15.0, 15.0)$ & $( 0.7, 2.0)$ & $( 5.0, 100.0)$ & $( 3.0,  50.0)$ \\
\cite{Meszarosova2014ApJ...788...44M}, Fig. 5, a1 &   1 & 62.549 & $(-20.0, 15.0)$ & $( 0.5, 1.5)$ & $(20.0, 200.0)$ & $( 2.0,  40.0)$ \\
\cite{Meszarosova2014ApJ...788...44M}, Fig. 5, a2 &   1 & 64.451 & $(-20.0, 15.0)$ & $( 0.5, 1.5)$ & $(15.0, 200.0)$ & $( 2.5,  40.0)$ \\
\cite{Meszarosova2014ApJ...788...44M}, Fig. 5, a3 &   1 & 59.221 & $(-15.0, 15.0)$ & $( 0.5, 1.5)$ & $(15.0, 200.0)$ & $( 2.5,  40.0)$ \\
\cite{Meszarosova2014ApJ...788...44M}, Fig. 5, b1 &   1 & 58.979 & $(-20.0, 10.0)$ & $( 0.5, 1.5)$ & $(20.0, 200.0)$ & $( 5.0,  40.0)$ \\
\cite{Meszarosova2014ApJ...788...44M}, Fig. 5, b2 &   1 & 57.616 & $(-20.0, 15.0)$ & $( 0.5, 1.5)$ & $(20.0, 200.0)$ & $( 3.0,  40.0)$ \\
\cite{Meszarosova2014ApJ...788...44M}, Fig. 5, b3 &   1 & 59.751 & $(-10.0, 20.0)$ & $( 0.5, 1.5)$ & $(20.0, 200.0)$ & $( 2.0,  40.0)$ \\
\cite{Meszarosova2014ApJ...788...44M}, Fig. 5, c2 &   1 & 66.454 & $(-20.0, 10.0)$ & $( 0.5, 1.5)$ & $(15.0, 150.0)$ & $( 1.5,  40.0)$ \\
\cite{Meszarosova2014ApJ...788...44M}, Fig. 5, c3 &   1 & 63.421 & $(-20.0, 10.0)$ & $( 0.5, 1.5)$ & $(10.0, 150.0)$ & $( 2.0,  40.0)$ \\
\cite{Meszarosova2014ApJ...788...44M}, Fig. 5, d2 &   1 & 67.289 & $(-20.0, 10.0)$ & $( 0.4, 1.3)$ & $(25.0, 150.0)$ & $( 2.0,  30.0)$ \\
\cite{Goddard2019A&A...624L...4G}, Fig. 3, 2    &   6 & 18.563 & $( -5.0, 35.0)$ & $( 0.7, 3.0)$ & $( 1.5,  20.0)$ & $(10.0, 200.0)$ \\
\cite{Goddard2019A&A...624L...4G}, Fig. 3, 3    &   6 & 22.939 & $( -5.0, 30.0)$ & $( 0.7, 3.0)$ & $( 1.5,  30.0)$ & $(10.0, 200.0)$ \\
\end{tabular}
\caption{Parameters of the random generation of QFP $Q(t)$. $W$ is weight that associated with probability of random sampling of the QFP type in the synthetic dataset. The higher weights are set up for more rare types of the $Q_0(t)$ like in \cite{Meszarosova2014ApJ...788...44M} and especially in \cite{Goddard2019A&A...624L...4G} in order to provide the higher diversity of the synthetic data set.
$t_{\text{start}}$ is uniformly distributed; $t_{\text{scale}}$, $\sigma_{\text{rise}}$, $\sigma_{\text{decay}}$ are log-uniformly distributed.}
\label{tab:qfp}
\end{table}

\subsection{Flare time profile pre-processing}\label{sec:preprocessing} 
Each time profile is preprocessed as described below.
Only the high-frequency component of the profile is passed into the neural network detector,
i.e. the time series after the subtraction of the slowly varying trend.
We tested three trend-removal approaches: the Savitzky–Golay polynomial filter (SavGol), Empirical Mode Decomposition (EMD), and Butterworth filter.
To define the slowly varying trend, we select the filter widths to be a half of the length of the synthetic light curve for the SavGol, and one sixth of the length for Butterworth filters. The Butterworth filter pass leads to a phase shift in the time series. 
Consequently, we apply this filter twice: first in the forward direction and then in the backward direction.
For EMD, we define the trend as the lowest-period modes.

The EMD approach is computationally expensive. Besides, it is difficult to automate because there is no a rigorous way to correspond an EMD mode (or modes) to a trend.
We have preferred the Butterworth filter for the trend removal because it works faster than other filters.
Each input time profile is normalized to its standard deviation.

\subsection{Neural network training and testing}
\begin{figure}[ht]
    \centering
    \includegraphics[width=\linewidth]{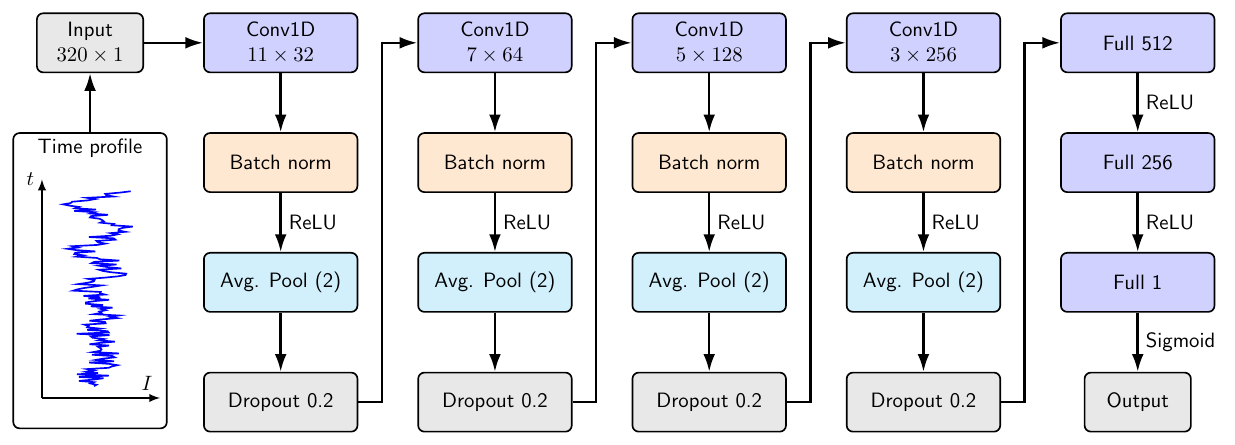}
    \caption{The convolutional neural network architecture used for the detector.
    Fully connected layers (Full on image) also have dropout rate $0.2$. For convolution block Conv1D $K \times C$, $K$ is kernel size and $C$ is channels count.}
    \label{fig:cnn-arch}
\end{figure}

We tested several neural-network architectures.
As a result, the detector is based on a CNN \cite{WangCNN2017}, see model details in Figure~\ref{fig:cnn-arch}.
We have not performed a hyper-parameters search and neural network optimization up to date.

The normalized to the own standard deviation, trend-free signal is passed into the network.
We used the standard split of the dataset into training, validation, and test sets (80:15:5). 
The network was trained as a binary classification with a loss function \texttt{BCEWithLogitsLoss} and \texttt{AdamW} optimizer.
The detector outputs the value ($p$) which then is rounded to a positive or negative trigger according to the chosen threshold $T$, $0 < T < 1$. The trigger is positive if $p > T$.

We tested the trained network detector on a synthetic test dataset and found that the computed standard binary classification metrics are too good: 99.8\% accuracy, 99.9\% precision and 99.7\% recall, for threshold $T = 0.5$.
These could indicate that the testing dataset is similar to the training dataset and validation dataset, i.~e., the diversity of the data set is not high enough.
We need to test the detector on a real data. Unfortunately, it is not possible to make a reliable statistics test because the number of the registered and published QFP-events in radio band with the known dates is very small. Therefore, we have reproduced the QFP wave train signatures in EUV band for the selected 16 published events \cite{Liu2011ApJ...736L..13L, Liu2012ApJ...753...52L, Zhou2022ApJ...930L...5Z, Miao2019ApJ...871L...2M, Qu2023ApJ...955...89Q, Ofman2018ApJ...860...54O, Zhou2024SCPMA..6759611Z, Nistico2014A&A...569A..12N, Shen2018ApJ...853....1S, Shen2022SoPh..297...20S, Zheng2018ApJ...858L...1Z, Miao2021ApJ...908L..37M, Sun2022ApJ...939L..18S, Li2024ScChE..67.1592L}
using SDO/AIA data.
Note that, in the EUV band, the QFPs are identified by different approach, particularly, by analysis of their dynamics in the running difference images.
Therefore, we decided to use the following approach.

For each of 16 events, the dime-distance map was constructed. The QFP wave train appears as a consequence of the slanted bright and dark stripes, and their inclination angle is defined by the phase speed of the wave train (see, for example, Figure~2 in \cite{Liu2011ApJ...736L..13L}). Then, interval of the distances was selected where the QFP wave train is the most visually pronounced. In this interval, we made several horizontal slices which are the averaged time profiles over three neighbour radio frequencies \cite{Liu2011ApJ...736L..13L}. We realize that these time profiles would be highly correlated, with a phase shift. However, as follows from the  results of numerical simulations, the shapes of the time profiles at different distances from the trigger would be different. Besides, the noises affect each time profile individually. So, we believe the time profiles independent if they were selected at some distance from each other.
\begin{figure}
    \centering
    \includegraphics[width=0.32\linewidth]{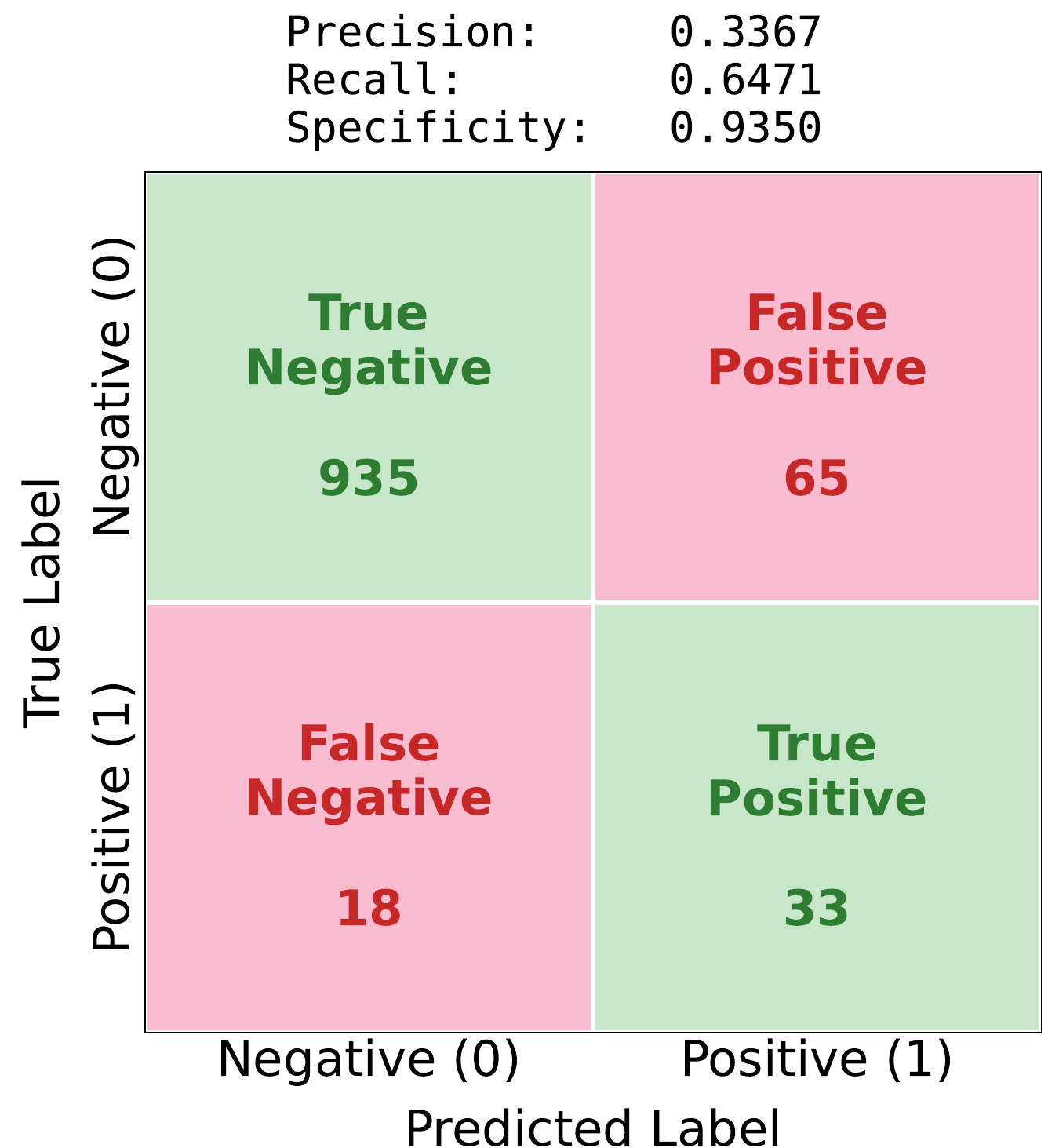}
    \includegraphics[width=0.32\linewidth]{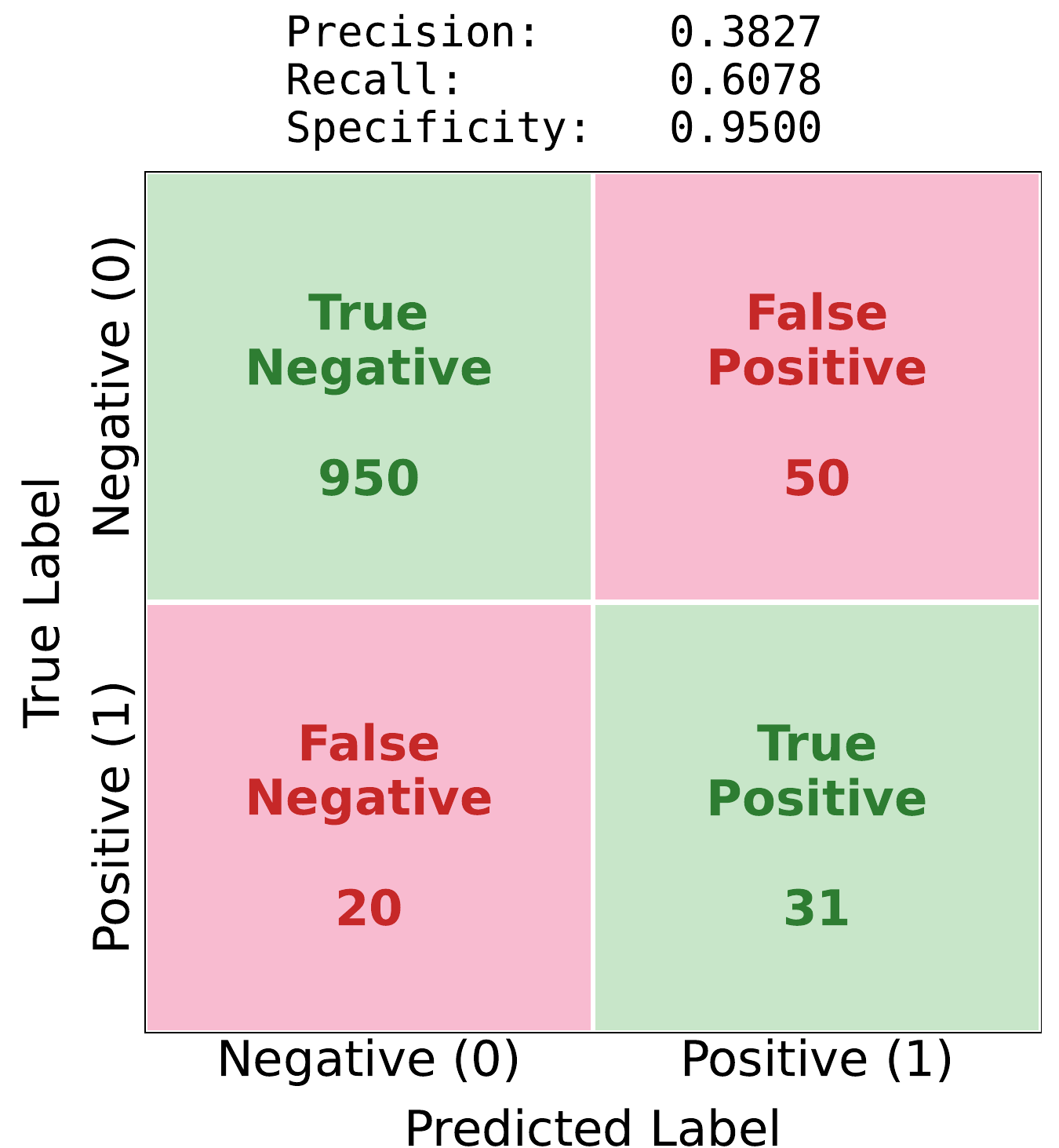}
    \includegraphics[width=0.32\linewidth]{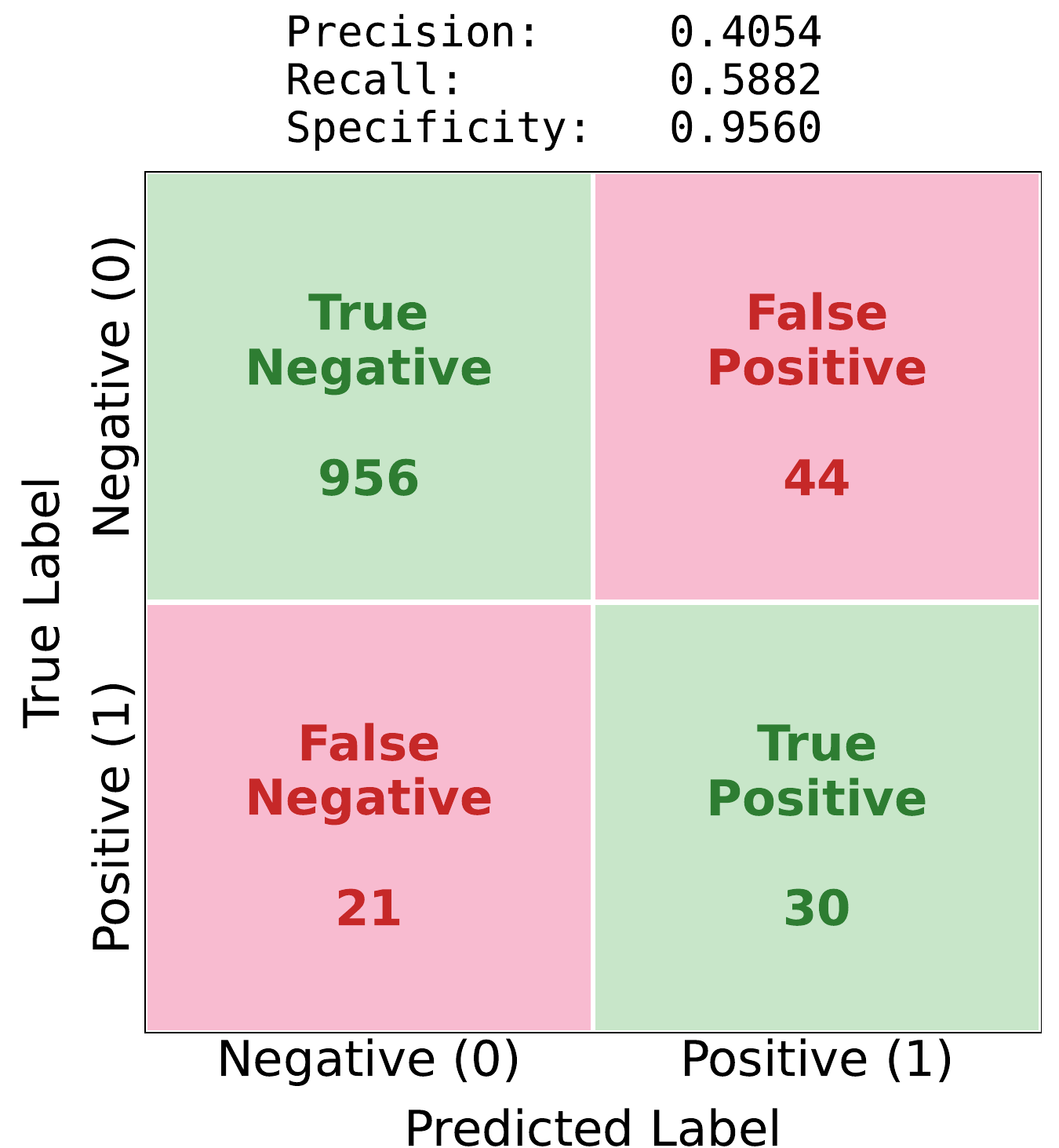}
    \caption{Confusion matrix for thresholds $T = 0.25$, $T = 0.5$, $T = 0.75$, from the left to the right panels.}
    \label{fig:confmat}
\end{figure}

We collect all the selected  time profiles 
(51 in total) 
to create the sample of real data to test our neural network.
Also we added
1000 "bad" time profiles which are
noises and spike profiles.
We obtained 38\% precision and 61\% recall and 95\% specificity for threshold $T = 0.5$ (see more in Figure~\ref{fig:confmat}).
Precision rate depends on ratio of negative to positive examples in the sample,
so, in out case with extra negative (similar to real), it can be low.
At this stage, it is not possible to provide definitive values for expected recall or false positive rates on real radio data. This is primarily due to the lack of a labeled reference set for real events that would allow for a statistically significant evaluation. 

We found a noticeable domain shift between results for the synthetic and real-data samples. We imply two main factors contributing to this shift. While synthetic profiles use modeled noise, the exact nature of instrumental and environmental noise in real-time radio observations is unknown. Besides, the physical transition from plasma density fluctuations to electromagnetic emission involves nonlinear effects that are difficult to replicate perfectly in a synthetic environment.

Note that the neural network has detected better those QFP-patterns which contain more oscillation cycles, in contrast with the signals with only a few oscillation cycles which rather look like individual consequent spikes or waves.
We will include the successful detections to retrain the neural network in the next step.

\section{Selection and processing of radio data} \label{sec:radiodata}
We used publicly available data from the Hiraiso radio
spectrograph (HiRAS) of the National Institute of Information and Communications
Technology (NICT) \cite{Kondo1997RCRL...43..231K}.
The spectrograph consists of three antennas operating in three frequency bands (20 -- 70\,MHz, 50 -- 500\,MHz, 500 -- 2500\,MHz).
The temporal resolution of the combined dynamic spectrum is $1 \text{ second}$.
It defines the lower bound of the periods of wave trains that can be detected, i.e.
$$
P > P_{\min}, \qquad P_{\min}=5 \text{ sec}.
$$

Since we do not know how widespread QFP wave trains are, we perform a targeted search for the dates on which the marker events were observed.

We consider the global coronal waves as the potential marker events based on the following reasons.
First, as it was mentioned in \cite{Liu2016AIPC.1720d0010L}, QFP wave trains often accompany the global coronal EUV waves. Both QFPs and global waves have a similar trigger,

for example, a solar flare, i.e. impulsive, localized energy releases.
If such a release is sufficiently powerful and compact to generate a global wave, it could also produce a QFP wave train. 
Second, type II radio bursts follow more than a half of the observed global waves \cite{Nitta2014SoPh..289.4589N}, so,  the global waves are pronounced in radio band.
And third, QFP-signatures are seen already in the radio band in several case studies \cite{Meszarosova2009ApJ...697L.108M, Kolotkov2018ApJ...861...33K, Karlicky2011A&A...529A..96K}

For definiteness, we began with the analysis of data for 2011 year, using the catalogue of the global waves (Large-scale Coronal Propagating Fronts catalogue, LCPFs catalogue) \cite{Nitta2013ApJ...776...58N}, its extended version is available on-line\footnote{\url{https://aia.lmsal.com/AIA_Waves/index.html}}.
For each date from the catalogue, the detector was applied to the entire observing time interval, 

without binding it to the exact time of the global wave.

Dynamic spectra for each selected date were pre-processed.
First, the trend was subtracted from the time series at each frequency (see Section~\ref{sec:method}
\ref{sec:preprocessing}).

The resulting high-frequency components formed a new dynamic spectrum,
which was then split into time windows of three different scales (320 sec, 640 sec, 1280 sec).

Three corresponding smoothing-filter widths were chosen ($\tau$ = 50 sec, $\tau$ = 100 sec, $\tau$ = 200 sec) for searching for different QFP-time scales; these define the upper bound of detectable periods:
$P \leq \tau$.
For windows (640 seconds and 1280 seconds), binning by two and four points, respectively, was performed so that the resulting implementation length equaled 320 counts, passed into the neural network.

The subsequent part of the processing of the dynamic spectra is based on the assumption that wave trains should be present in some frequency band.
The width of this band depends on the wave train mechanism and emission mechanism but the cross-correlation coefficients between time profiles on adjacent frequencies should be high, even when the QFPs have a phase shift at the neighbouring radio frequencies \cite{Kuznetsov2006SoPh..237..153K, Karlicky2013A&A...550A...1K}.
From this consideration, within each time window, for all radio frequencies, the
cross-correlation coefficients were computed between the time profile on the $i$-th
frequency and the one on the $(i+k)$-th frequency, with $k$ is integer index
where $k$ is non-zero integer with absolute values $\le 10$.
The mean value over 20 obtained cross-correlation coefficients forms the new dynamic spectrum as metric for the next detection step.

This step decreases significantly the noise impact, limits the possible area of the wave train search, and optimizes the computing time for the detector.

\begin{figure}[ht]
    \centering
    \includegraphics[width=\linewidth]{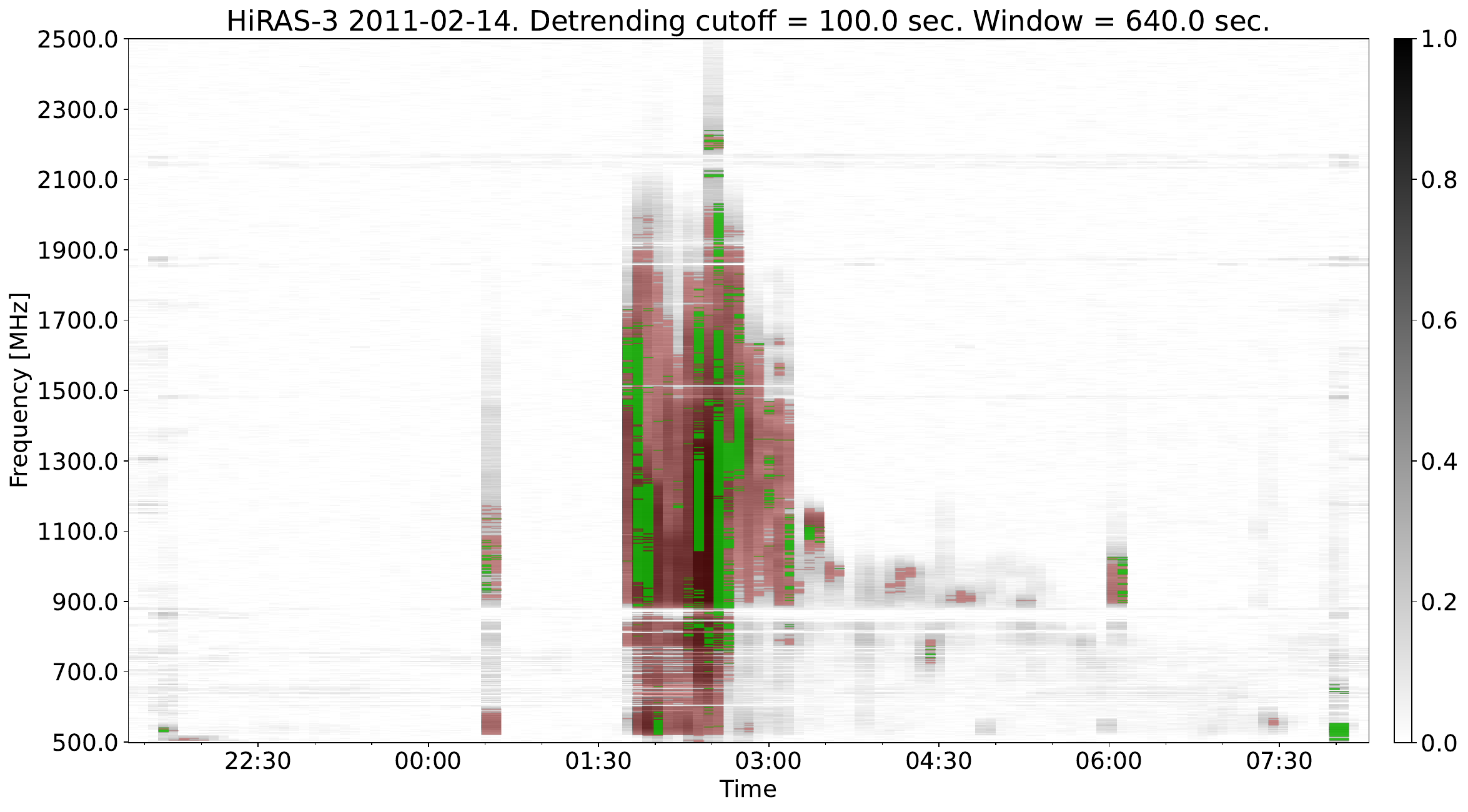}
    \caption{The dynamic spectrum presenting the mean cross-correlation coefficients (white-black). 
    Regions where the mean cross-correlation coefficients exceed 0.25 are highlighted with gradation of the transparent red. The time profiles in these regions were passed as an input into the neural network detector. 
    Areas with the positive detector results (for threshold $T = 0.5$) are highlighted with green.
    }
    \label{fig:cross_corr}
\end{figure}

An example of the dynamic spectrum containing both the mean cross-correlation coefficients and the results from the detector are shown in Figure~\ref{fig:cross_corr}.
Quasi-periodic events identified in this way are reported as the 
QFP-candidates.

We should note that frequency bins and threshold were selected empirically. 
If we consider a small bin sizes ($\lesssim 4$) for HiRAS radio frequency then we obtain too many false positive detections (around a half of dynamic spectrum when true positive is around a few percent) because of both noise correlation at neighbour radio frequencies and narrow-band spikes. The low threshold (< 0.2) also leads to an extra false positive rate (around a third of the dynamic spectrum) while the high threshold (> 0.4) leads to extremely high false negatives (the recall estimate less than 10\%). Specific values of the bins may depend on a spectral resolution of an  instrument. For example, for the lower spectral resolution, probably, a lower bin size could be enough.

\section{Results}\label{sec:results}
In the LCPFs catalogue for the year 2011, we found 102 global waves,
however only 50 of them were available in HiRAS data.
We performed the search 

for each of those dates and have detected 
132 QFP wave trains candidates using the described-above methods.

\begin{figure}
    \centering
    \includegraphics[width=0.49\linewidth]{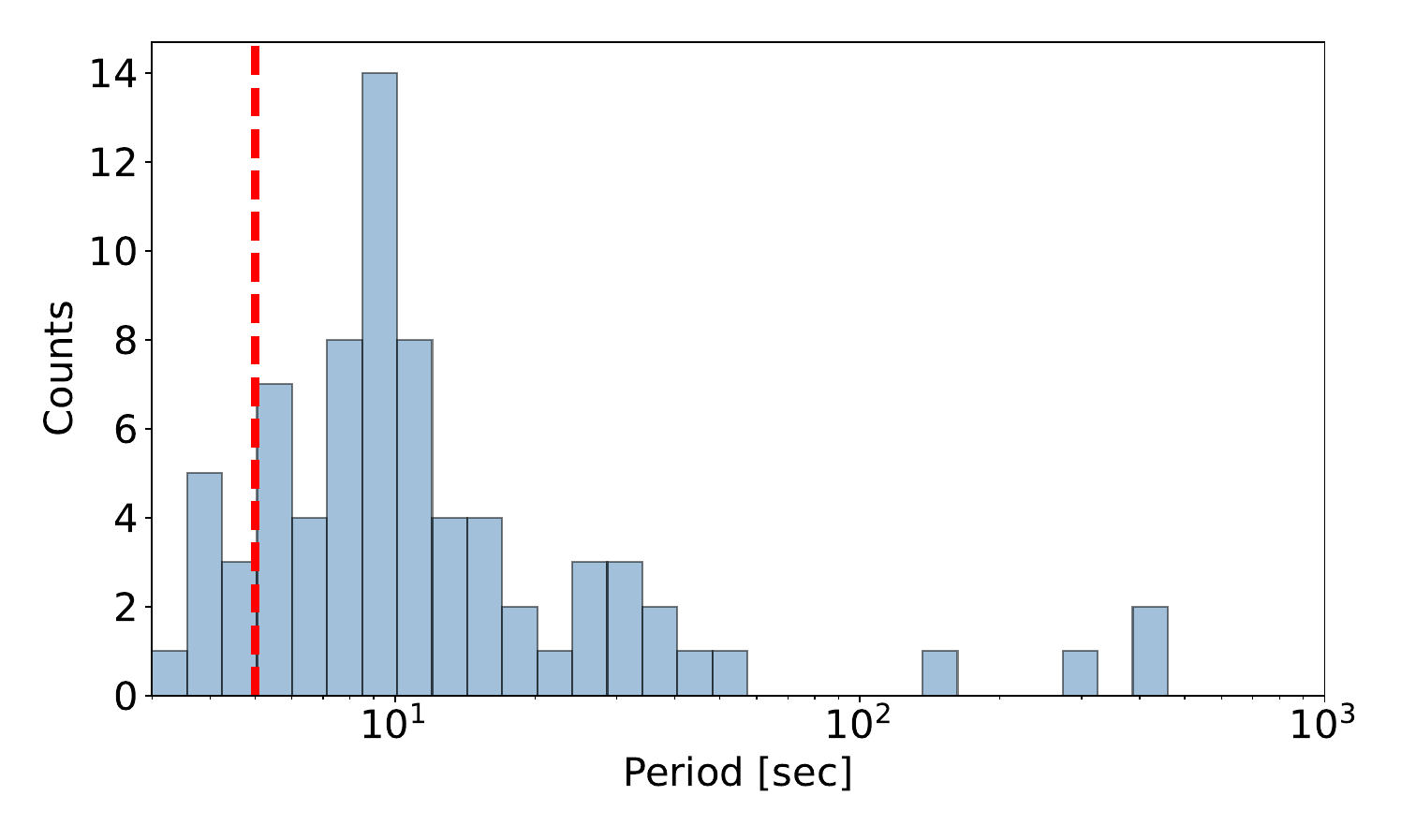}
    \includegraphics[width=0.49\linewidth]{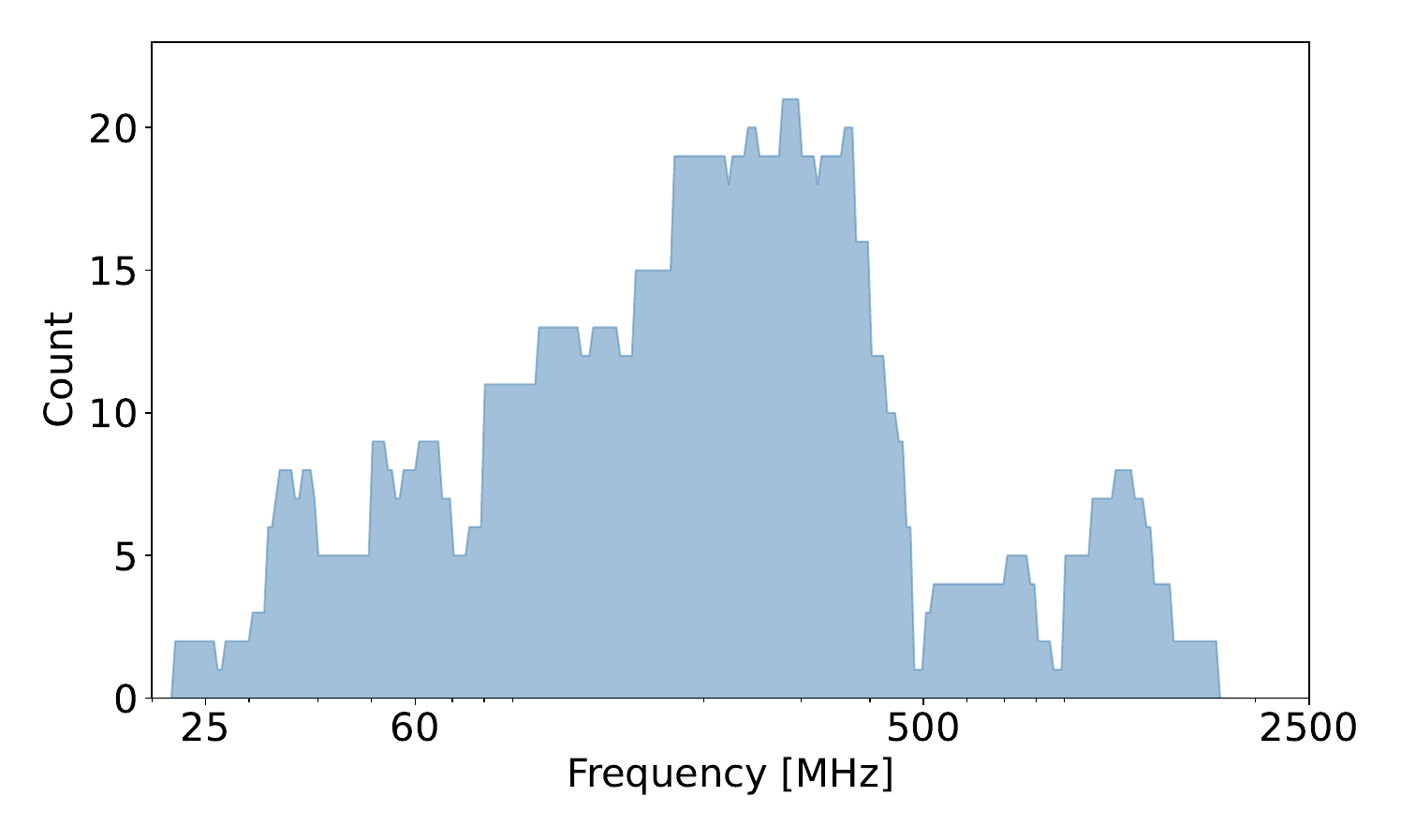}
    \caption{Distribution of parameters for QFP-candidates: average periods (left panel) and radio frequencies where they have been detected}.
    \label{fig:stats}
\end{figure}

The great majority of the detections (83 detections) is found in HiRAS2 data vs 31 detections in HiRAS1 and 18 detections in HiRAS3, with the following ratios positive/negative detections: 15/18 for HiRAS1, 43/41 for HiRAS2, 14/4 for HiRAS3. 
Distribution of the positive detections over radio frequencies is shown in the right panel in Figure~\ref{fig:stats}.
Note that a high degree of the cross-correlation or even a (false) positive neural detector result can also often be an artifact of the smoothing of a single short high-amplitude spike in the flare time profile.

\begin{figure}[ht]
  \centering
  \includegraphics[width=\linewidth]{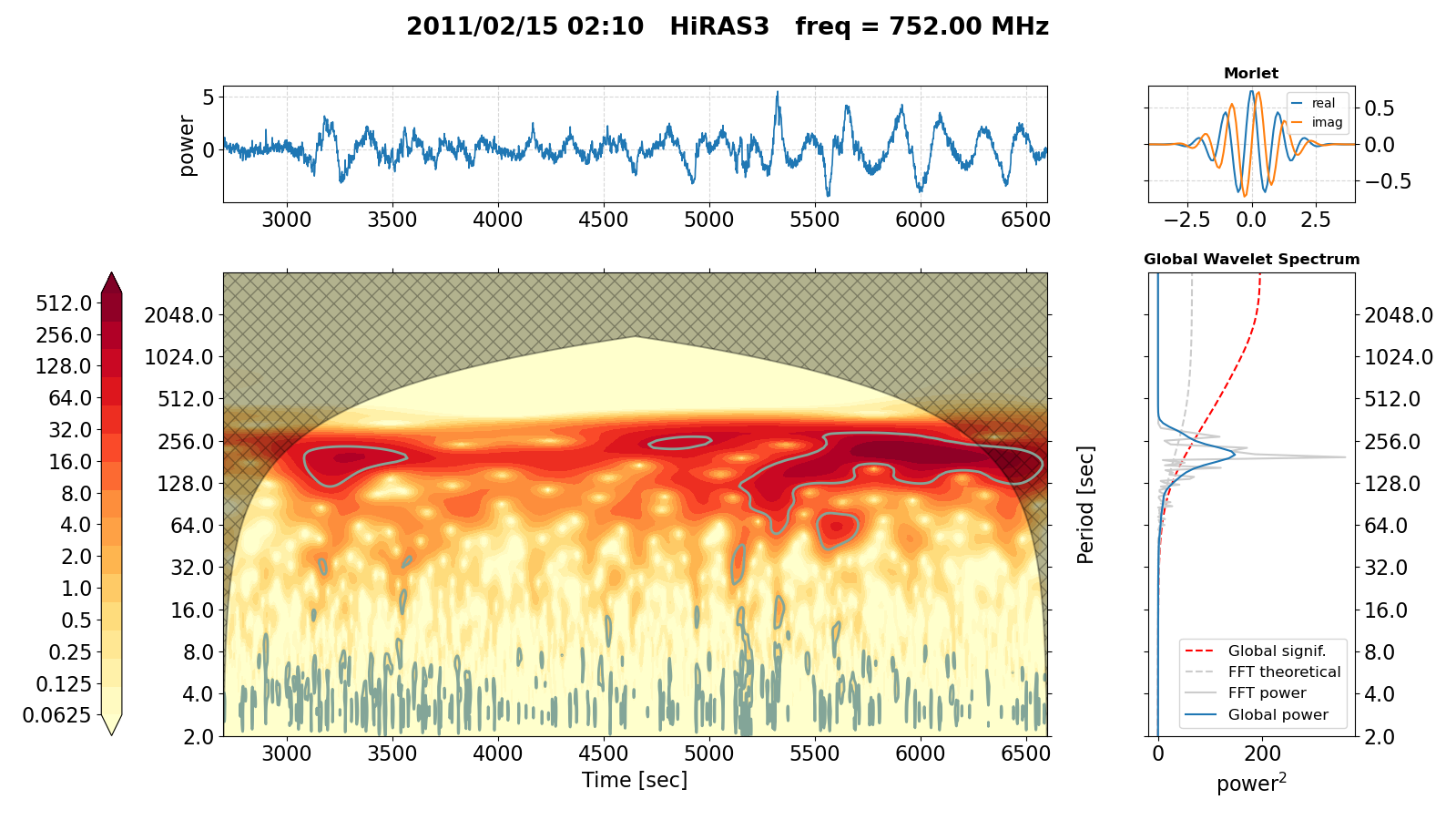}
  \caption{Time profile (upper-left panel) and its wavelet  spectrum (lower-left panel) for two QFP-candidates. 
  Upper-right panel presents the corresponding wavelet mother function.
  The time profile was obtained by averaging over three neighbour radio frequencies. 
  The cutoff period at the trend-removal stage is 300 seconds. 
  The global wavelet spectrum in shown in lower-right panel. }
  \label{fig:wavelet_ex}
\end{figure}

\begin{figure}[ht]
  \centering
  \includegraphics[width=\linewidth]{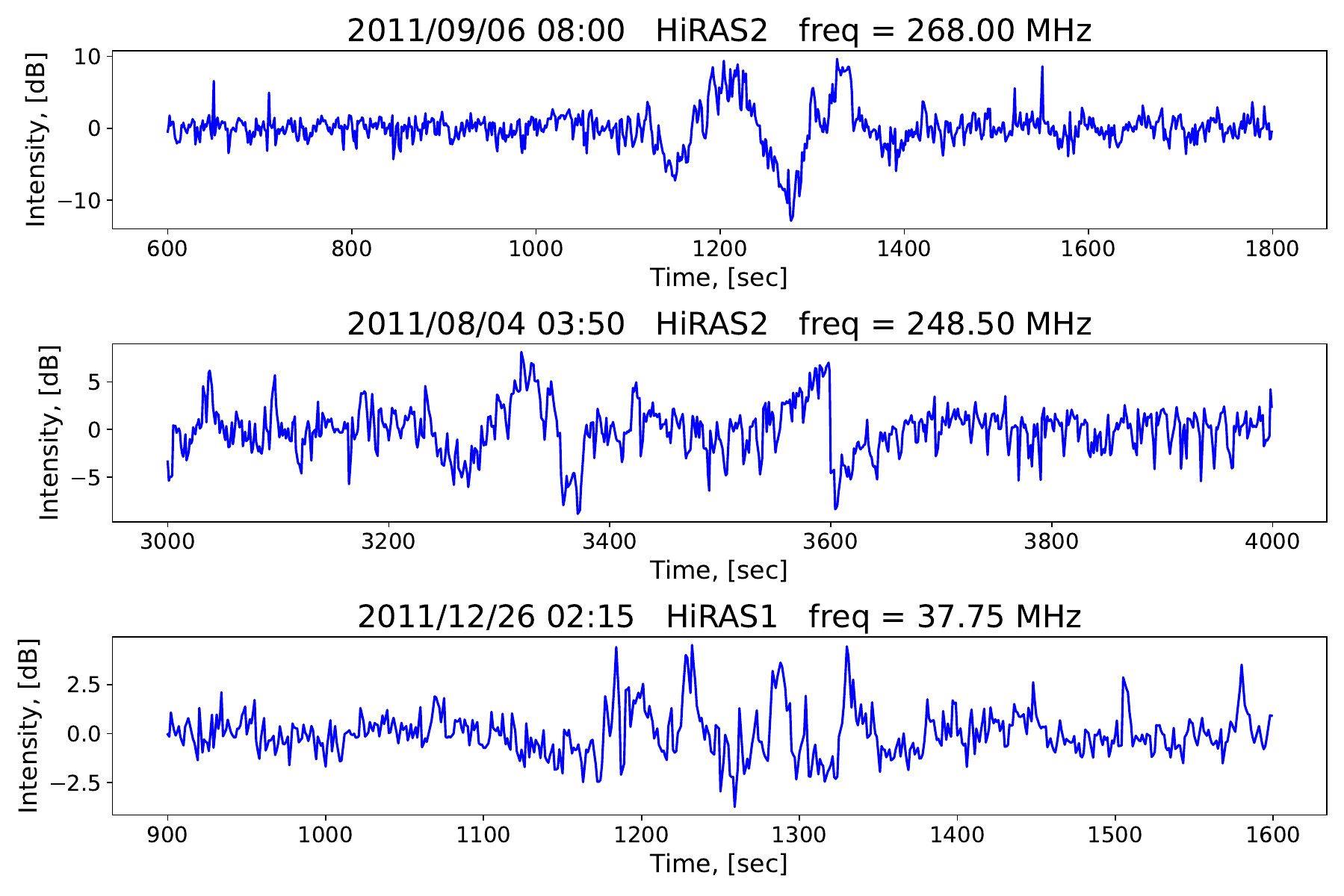}
  \caption{
  Examples of the detrended time profiles for three QFP-candidates. Each time profile was obtained by averaging over three neighbour radio frequencies. Cutoff period at the trend-removal stage is 200 seconds.}
  \label{fig:exs}
\end{figure}

Our automatic detector revealed 75 candidates phenomenologically similar to a QFP wave train.
For all the identified candidates, we plot the flare time profiles, their trends and high‑frequency components, as well as wavelet spectra of the high‑frequency component for additional visual control (see an example in Figure~\ref{fig:wavelet_ex}, also see another examples in Figure~\ref{fig:exs}).
The full list of the detected wave trains and their frequencies is presented in Table~\ref{tab:results}.
Among them, 26 detections are found to happen within 20 minutes after the global coronal wave and 20 detections have no connection with the global waves. There are no wave trains detected exactly before a global wave.
The other 26 events are questionable because they are phenomenologically similar to the spikes and possibly are instrumental or processing artifacts.

However, at this stage while the number of registered wave train events still being low, we prefer a higher recall score and accept any sentient precision score.
In another words, currently, we prefer to detect something false than to skip something true. 
Therefore we also put these detections to Table~\ref{tab:results} minding that all the detected wave trains need in an additional analysis.

Note that the instrumental noisy narrowband stripes at some radio frequencies could split a broadband QFP to several different parts. In this case, we  see several detection of one QFP-event. Accounting for this feature, we found that only 50 candidates of 75 detections are independent events, with 13 candidates connected with the global waves. Table~\ref{tab:results} includes  all the detections but not the independent events.

Histogram for the estimated periods of the detected QFP candidates id shown in the left panel in Figure~\ref{fig:stats}.
In practice, we have detected mostly QFPs with 3--4 cycles of oscillation, and half of them are questionable due to high noise. Confident detection of QFPs candidates with 1--2 cycles of oscillation is extremely difficult, as it is similar to a spike signal. Additionally, we note that QFP candidates with periods $\lesssim 5$ seconds may be unreliable, as they contain too few counts per period.

\begin{table}[ht]
\centering
\begin{tabular}{c|l||c|l}
Date & Freq. range, MHz & Date & Freq. range, MHz \\ \hline
2011-01-13T06:30 & HiRAS2: $  175-300$ & 2011-08-02T06:25 & HiRAS3: $1100-1250$ \\
2011-01-26T22:55 & HiRAS2: $  250-460$ & 2011-08-04T03:50 & HiRAS2: $  150-220$ \\
2011-02-12T01:25 & HiRAS2: $  200-475$ & 2011-08-04T03:55 & HiRAS1: $    57-67$ \\
2011-02-12T04:45 & HiRAS2: $  200-475$ & 2011-08-04T03:55 & HiRAS1: $    40-50$ \\
2011-02-12T04:55 & HiRAS2: $  200-475$ & 2011-08-04T03:55 & HiRAS1: $    27-40$ \\
2011-02-14T01:45 & HiRAS2: $  275-400$ & 2011-08-04T04:30 & HiRAS3: $  600-800$ \\
2011-02-14T02:45 & HiRAS2: $  150-475$ & 2011-08-09T08:00 & HiRAS2: $  125-450$ \\
2011-02-14T03:40 & HiRAS2: $  275-400$ & 2011-08-09T08:00 & HiRAS3: $  500-800$ \\
2011-02-14T04:40 & HiRAS2: $  150-475$ & 2011-08-09T08:05 & HiRAS1: $    40-70$ \\
2011-02-14T05:45 & HiRAS1: $    32-40$ & 2011-08-09T08:15 & HiRAS3: $ 900-1400$ \\
2011-02-14T05:50 & HiRAS3: $1000-1200$ & 2011-09-06T01:45 & HiRAS1: $    60-67$ \\
2011-02-14T23:40 & HiRAS2: $  225-350$ & 2011-09-06T01:40 & HiRAS2: $  100-250$ \\
2011-02-15T00:35 & HiRAS2: $  275-375$ & 2011-09-06T01:55 & HiRAS1: $    22-26$ \\
2011-02-15T02:10 & HiRAS3: $ 900-1700$ & 2011-09-06T22:15 & HiRAS2: $  275-425$ \\
2011-02-15T02:15 & HiRAS2: $  250-425$ & 2011-09-06T22:15 & HiRAS2: $  175-300$ \\
2011-02-15T03:05 & HiRAS1: $    46-53$ & 2011-09-06T22:15 & HiRAS2: $  100-275$ \\
2011-02-15T03:10 & HiRAS2: $   80-150$ & 2011-09-06T22:25 & HiRAS1: $    22-27$ \\
2011-02-15T04:25 & HiRAS1: $    34-40$ & 2011-09-06T22:15 & HiRAS3: $  500-850$ \\
2011-02-15T04:30 & HiRAS2: $   50-250$ & 2011-09-06T22:15 & HiRAS3: $ 900-1700$ \\
2011-02-15T04:35 & HiRAS1: $    30-40$ & 2011-09-07T22:40 & HiRAS2: $  150-225$ \\
2011-02-17T23:50 & HiRAS1: $    33-35$ & 2011-09-07T22:40 & HiRAS2: $  325-400$ \\
2011-02-18T01:10 & HiRAS2: $  240-320$ & 2011-09-22T01:10 & HiRAS1: $    32-70$ \\
2011-02-18T01:20 & HiRAS2: $   80-120$ & 2011-09-22T04:25 & HiRAS2: $  175-275$ \\
2011-02-18T05:00 & HiRAS2: $  225-275$ & 2011-09-24T00:25 & HiRAS3: $  700-770$ \\
2011-02-18T05:00 & HiRAS2: $   80-140$ & 2011-09-24T04:45 & HiRAS1: $    32-36$ \\
2011-02-24T07:30 & HiRAS3: $ 900-1300$ & 2011-09-25T02:30 & HiRAS3: $  520-600$ \\
2011-02-24T07:40 & HiRAS1: $    27-55$ & 2011-09-25T05:40 & HiRAS3: $1000-1400$ \\
2011-03-08T03:15 & HiRAS2: $  250-460$ & 2011-10-01T00:40 & HiRAS1: $    37-46$ \\
2011-03-08T06:35 & HiRAS2: $   80-225$ & 2011-10-02T00:45 & HiRAS2: $   50-275$ \\
2011-03-09T03:40 & HiRAS2: $  360-460$ & 2011-12-12T21:55 & HiRAS3: $ 650-1300$ \\
2011-03-09T06:30 & HiRAS2: $  100-250$ & 2011-12-13T04:15 & HiRAS2: $   80-200$ \\
2011-05-16T00:20 & HiRAS2: $  275-375$ & 2011-12-21T23:30 & HiRAS2: $   50-250$ \\
2011-06-02T07:30 & HiRAS2: $   75-100$ & 2011-12-22T01:00 & HiRAS2: $  175-200$ \\
2011-06-04T08:30 & HiRAS2: $  225-275$ & 2011-12-26T02:15 & HiRAS1: $    35-39$ \\
2011-06-04T08:30 & HiRAS2: $  350-400$ & 2011-12-26T02:15 & HiRAS2: $   50-200$ \\
2011-07-30T01:10 & HiRAS2: $  275-375$ & 2011-12-26T02:15 & HiRAS3: $  530-650$ \\
2011-07-30T03:10 & HiRAS2: $  275-375$ & 2011-12-27T01:40 & HiRAS2: $  150-530$ \\
2011-08-01T23:35 & HiRAS2: $   50-150$ &                  &                     \\
\end{tabular}
\caption{Found QFP candidates.} 
\label{tab:results}
\end{table}

\section{Discussion and conclusion}\label{sec:discussion}
We performed an automatic search for the fast wave trains in radio data over year 2011 using the classifying neural network/machine learning methods. We analyzed the HiRAS dynamic radio spectra within the 20 MHz -- 2.5 GHz frequency band.  We considered 50 global coronal EUV waves as marker events for a more targeted search, and found that 13 global waves had the QFP-response in the radio emission at different frequencies. 
We have estimated binomial confidence intervals at the 95\% level using the Wilson method and obtained the value 26\% (with a confidence interval of 16\%--39\%) for our results for the radio band and the value 32\% (with a confidence interval of 29\%--39\%) for the EUV band \cite{Liu2016AIPC.1720d0010L}. 

In our study, we conducted a blind search for QFP-candidates throughout the entire observation period, not just around the time when a global wave occurred. We don't know when the QFP-candidate will appear, nor do we know the specific distribution of QFPs over time. Therefore, we assume this distribution to be uniform.
Taking into account the duration of the HiRAS observation period and the maximal time lag between a global wave and he corresponding QFP-candidate, we estimate the probability of a random coincidence between the QFP and a global event to be less than 10\%. This estimate is significantly lower than our obtained value of 26\%. Additionally, the Z-score is 4.0, i.e. rather high.

There are two known global scenarios of how a fast wave train could  affect the emission in the radio band: modulation of the 
parameters of the emitting plasma, and the modulation of the acceleration rate and spectrum of non-thermal electrons.
In the first scenario, a wave train modulates plasma parameters such as the plasma density and magnetic field and, 

depending on the radio frequency band in which the wave train is observed, it modulates either the intensity of the gyro-synchrotron emission or the coherent plasma emission.
For example, it was found that the modulation of the intensity in decimetre type-IV bursts is highly correlated (correlation coefficient $>0.5$) over a wide frequency range (1.2 -- 4.5 GHz), the higher-frequency part of which corresponds to the gyrosynchrotron mechanism \cite{Meszarosova2009ApJ...697L.108M}.
By relating the start and end times of the wave train and assuming a particular Alfv{\'e}n speed in the waveguide,
the authors estimated the distance from the detection point to the trigger and the half-width of the coronal loop acting as a waveguide (i.~e., the transverse scale of the plasma density inhomogeneity).
In a lower-frequency band ($\lesssim 2$ GHz), where the coherent plasma mechanism dominates,
modulations of the plasma frequency produce a variety of fine structures in the dynamic spectra.

Examples include zebra-stripes modulations associated with harmonics of the upper hybrid frquency \cite{Karlicky2013A&A...550A...1K}, wave-like "fiber"{} bursts \cite{Kuznetsov2006SoPh..237..153K,Meszarosova2009A&A...502L..13M,Karlicky2013A&A...550A...1K}, and fine structures of type-III bursts \cite{Kolotkov2018ApJ...861...33K}.

In the second scenario, the wave train modulates the production of the accelerated electrons. For example, it could propagate through a current sheet and regulates the turbulent process of magnetic island (plasmoid) formation \cite{Barta2001A&A...379.1045B}.
In \cite{Karlicky2011A&A...529A..96K}, this mechanism was used to interpret the temporal structure of type-IV decimetre bursts (0.8 -- 2 GHz) as wave trains of narrowband spikes.
Another possibility is that the fast wave train could interact with the shock front of a coronal mass ejection leading to acceleration of  electrons, producing the bump-on-tail instability, and subsequently emitting the radio waves at the local electron plasma frequency (type-II radio burst). Such event was found preceding the type II radio burst \cite{Goddard2016A&A...594A..96G}. Besides, wave trains may, potentially, modulate the electron acceleration process, for example, by modulating the coalescence instability \cite{Tajima1987ApJ...321.1031T, Kolotkov2016PhRvE..93e3205K}, imparting a characteristic temporal structure to type-III bursts.

In our study, it is not possible to detect the first-scenario QFPs neither as registering the fine frequency drifts of the coherent plasma emission pronounced at the sub-second time scales as the HiRAS time cadence is 1 second, nor as the modulation of the gyrosynchrorton emission as that usually occurs at frequencies higher than the HiRAS frequency band. So we connect our results with the second scenario when the QFP wave train modulates the electron acceleration process, i.~e. the type-II or type-III radio bursts. 
The overall question remaining open is how a wave train can reach the corresponding altitudes in the solar atmosphere. 
This indicates the need for dedicated numerical simulations are highly required.

All QFP-candidates detected in this study require an independent validation, for example, with multi-instrument analysis.
Future work will involve expanding the processed data to include 

multi-wavelength observations (EUV, HXR) and multi-instruments observations, and refining the detector approach.

\dataccess{The HiRAS radio data used in this study are Courtesy of National Institute of Information and Communications Technology (NICT). They are publicly available at \url{https://solarobs.nict.go.jp/}.
The SDO data are courtesy of NASA/SDO/AIA science team, publicly available at the Joint Science Operations Center (JSOC).
The catalogue of the Large-scale Coronal Propagating Fronts is publicly available on-line \url{https://aia.lmsal.com/AIA_Waves/index.html}}.

\funding{This study is supported  under the State Assignments Nos. 125020501551-5 and FFUG-2024-0002.}

\conflict{The authors declare no conflict of interest. }

\bibliographystyle{RS}

\end{document}